\documentclass[12pt]{article}
\usepackage{amssymb,amsmath,amscd}
\usepackage{color}

\newcommand{\sw}{\stackrel{\star}{\wedge}}

\begin{document}

\topmargin -2pt


\headheight 0pt


\vspace{5mm}

\begin{center}
{\Large \bf Noncommutative BTZ Black Hole \\
in Different Coordinates\footnote{Talk given at CosPA2008, October 2008.}} \\

\vspace{10mm}

{\sc Ee Chang-Young}
\\

\vspace{1mm}

{\it Department of Physics, Sejong University, Seoul 143-747, Korea}
\\

\vspace{5mm}

cylee@sejong.ac.kr \\

\vspace{10mm}
{\bf ABSTRACT}
\end{center}


\noindent
\noindent

We consider noncommutative BTZ black hole solutions in two different
coordinate systems, the polar and rectangular coordinates.
The analysis is carried out by obtaining noncommutative solutions of
$U(1,1)\times U(1,1)$ Chern-Simons theory on $AdS_3$ in the
two coordinate systems via the Seiberg-Witten map.
This is based on the noncommutative extension of the equivalence
between the classical BTZ solution and
the solution of ordinary $SU(1,1)\times SU(1,1)$
Chern-Simons theory on $AdS_3$.
The obtained solutions in these noncommutative
coordinate systems become different
in the first order of the noncommutativity parameter $\theta$.
\\




%

\section{Introduction}

 The most used canonical  commutation relation for noncommutative spacetime
  is modeled on the commutation relation of quantum mechanics:
\begin{equation}
\label{moyal_nc}
[\hat{x}^\alpha,\hat{x}^\beta]=i \theta^{\alpha\beta},
\end{equation}
where $\theta^{\alpha\beta}=-\theta^{\beta\alpha}$ are constants.
It has been also known that a theory on a deformed spacetime with the above
canonical relation is equivalent to a theory
on commutative spacetime in which any product of functions of noncommutative coordinates
are replaced with a deformed $\star$-product of the same functions of commutative coordinates,
the so-called Moyal product \cite{gromoyal} which is defined by
\begin{equation}
\label{moyalprd}
(f\star g)(x)\equiv \left.\exp\left[\frac{i}{2}\theta^{\alpha\beta}\frac{\partial}
{\partial x^{\alpha} }\frac{\partial}{\partial y^{\beta} }\right] f(x)g(y)\right|_{x=y} .
\end{equation}

Using the Moyal product many works on noncommutative spacetime have been carried out and
especially in \cite{sw99} a map between a gauge theory on noncommutative spacetime and
one on commutative spacetime, the so-called Seiberg-Witten map, was established.

So far, we are accustomed to take general covariance for granted.
 General covariance in ``a noncommutative space\footnote{
We only deal with space-space noncommutativity here, and
 we use (noncommutative) space and (noncommutative) spacetime interchangeably.}" would mean the
equivalence among different coordinate systems in this noncommutative space.
However, different coordinate systems in ``a noncommutative space"
generally have different commutation relations which are not exactly equivalent to each other.
Therefore, if we work with different coordinate systems in ``a noncommutative space",
we may end up with different results.
If this happens, it would contradict our usual notion of general covariance.
Seiberg \cite{ns2005} has already pointed out that general covariance would be broken
in  theories with emergent spacetime  among which model theories on noncommutative spaces are
 included.
As a step on this issue,
here we investigate the noncommutative BTZ solutions
in the polar and rectangular coordinate systems.
We carry this by obtaining the solutions of
 $U(1,1)\times U(1,1)$ Chern-Simons theory on noncommutative $AdS_3$ in the
two coordinate systems via the Seiberg-Witten map.
%


\section{Noncommutative Chern-Simons gravity}

The action of the $(2+1)$ dimensional noncommutative $U(1,1)\times U(1,1)$
 Chern-Simons theory with the negative
cosmological constant $\Lambda=-1/l^2$ is given by up to boundary terms \cite{bcgss,ckmz},
\begin{eqnarray}
\label{action}
&&\hat{S}(\mathcal{\hat{A}}^{+},\mathcal{\hat{A}}^{-})=
\hat{S}_{+}(\mathcal{\hat{A}}^{+})-\hat{S}_{-}(\mathcal{\hat{A}}^{-}), \\
&& \hat{S}_{\pm}(\mathcal{\hat{A}}^{\pm})=
\beta\int \rm Tr(\mathcal{\hat{A}}^{\pm} \sw d\mathcal{\hat{A}}^{\pm}+\frac{2}{3}
\mathcal{\hat{A}}^{\pm}\sw \mathcal{\hat{A}}^{\pm} \sw \mathcal{\hat{A}}^{\pm}),\nonumber
\end{eqnarray}
where $\beta=l/16\pi G_{N}$ and  $G_{N}$ is the three dimensional Newton constant.
Here
$
\mathcal{\hat{A}^{\pm}}=\mathcal{\hat{A}}^{A\pm}\tau_{A}
=\hat{A}^{a\pm}\tau_{a}+\hat{B}^{\pm}\tau_{3},
$
with  $A=0,1,2,3$,  $~ a={0,1,2},$ ~ $\mathcal{\hat{A}}^{a\pm}=\hat{A}^{a\pm}$,
 $~ \mathcal{\hat{A}}^{3\pm}=\hat{B}^{\pm}$,
 and the deformed wedge product $\sw$ denotes  that
$
A \sw B \equiv A_{\mu} \star B_{\nu}~dx^{\mu} \wedge dx^{\nu}.
$
The noncommutative $SU(1,1) \times SU(1,1)$ gauge fields $\hat{A}$ are expressed in terms of the triad
$\hat{e}$ and the spin connection $\hat{\omega}$ as %
$\label{nc_cs3grav}
 \hat{A}^{a\pm}:=\hat{\omega}^{a}\pm \hat{e}^{a}/{l}.
$
In terms of $\hat{e}$ and  $\hat{\omega}$
the action becomes \cite{ckmz}
\begin{eqnarray}
\label{reaction}
\hat{S}\!\!&=&\!\!\frac{1}{8\pi G_{N}}\int\left(\hat{e}^{a}\sw \hat{R}_{a}
+\frac{1}{6l^2}\epsilon_{abc}\hat{e}^{a}\sw\hat{e}^{b}
\sw\hat{e}^{c}\right)
\nonumber \\
\!\!&-&\!\! \frac{\beta}{2}
\int\left(\hat{B}^{+}\sw d\hat{B}^{+}
+\frac{i}{3}\hat{B}^{+}\sw\hat{B}^{+}
\sw\hat{B}^{+}\right)
+\frac{\beta}{2}
\int\left(\hat{B}^{-}\sw d\hat{B}^{-}
+\frac{i}{3}\hat{B}^{-}\sw \hat{B}^{-}
\sw\hat{B}^{-}\right)
\nonumber \\
&+&\frac{i\beta}{2} \int( \hat{B}^{+}-\hat{B}^{-})\sw
\left(\hat{\omega}^{a}\sw \hat{\omega}_{a}+\frac{1}{l^2}\hat{e}^{a}
\sw\hat{e}_{a}\right)
\nonumber \\
&+&\frac{i\beta}{2l}\int( \hat{B}^{+}+\hat{B}^{-})\sw
\left(\hat{\omega}^{a}\sw \hat{e}_{a}+\hat{e}^{a}
\sw \hat{\omega}_{a}\right),
\end{eqnarray}
up to surface terms, where  $\hat{R}^{a}=d\hat{\omega}^{a}
+\frac{1}{2}\epsilon^{abc}\hat{\omega}_{b}\stackrel{\star}{\wedge}\hat{\omega}_{c}$.
The equation of motion can be written as follows.
\begin{eqnarray}
\label{nccurtensor}
\hat{\mathcal{F}}^{\pm} \equiv d\hat{\mathcal{A}}^{\pm}+ \hat{\mathcal{A}}^{\pm}
\sw\hat{\mathcal{A}}^{\pm}=0.
\end{eqnarray}
%
In the commutative limit this becomes,
\begin{eqnarray}
\label{ccurtensor}
F^{\pm} \equiv d A^{\pm}+ A^{\pm}\wedge A^{\pm}=0, ~~ d B^{\pm}= 0,
\end{eqnarray}
and the first one can be rewritten as
\begin{equation}
R^{a} + \frac{1}{2l^2}\epsilon^{abc}e_{b}\wedge e_{c}=0, ~~
T^{a} \equiv de^{a}+\epsilon^{abc}\omega_{b}\wedge e_{c}= 0.
\end{equation}
The solution of the decoupled EOM for $SU(1,1)\times SU(1,1)$ part
was obtained in \cite{cgm}:
\begin{eqnarray}
\label{triad}
e^{0}&=& m\left(\frac{r_{+}}{l}dt-r_{-}d\phi\right),~
e^{1}=\frac{l}{n}dm,~
e^{2}=n\left(r_{+}d\phi-\frac{r_{-}}{l}dt\right), \nonumber
\\
\label{spinc}
\omega^{0}&=& -\frac{m}{l}\left(r_{+}d\phi-\frac{r_{-}}{l}\right),~
\omega^{1}=0,~~~~~
\omega^{2}=-\frac{n}{l}\left( \frac{r_{+}}{l}dt-r_{-}d\phi\right),
\end{eqnarray}
where $m^2=(r^2-r_{+}^2)/(r_{+}^2-r_{-}^2)$,~ $n^2=(r^2-r_{-}^2)/(r_{+}^2-r_{-}^2)$,
and $r_+,~ r_-$ are the outer and inner horizons respectively.
There it was also shown to be equivalent to the ordinary BTZ black hole solution \cite{btz}:
\begin{equation}
ds^2=-N^2dt^2+N^{-2}dr^2+r^2(d\phi+N^{\phi}dt)^2,
\end{equation}
where $N^2=(r^2-r_{+}^2)(r^2-r_{-}^2)/l^2 r^2$ and $N^{\phi}=-r_{+}r_{-}/lr^2$.

\section{Noncommutative BTZ solution in polar coordinates}

%
Based on the above noncommutative extension of the equivalence between  3D gravity and
$SU(1,1)\times SU(1,1)$ Chern-Simons theory, we now get the noncommutative
BTZ solution in the polar coordinates following \cite{ell01} using the Seiberg-Witten map.
The Seiberg-Witten map which dictates the following equivalence relation between
ordinary and noncommutative gauge transformations \cite{sw99},
\begin{eqnarray}
\label{swe}
\hat{\mathcal{A}}_{\mu}(\mathcal{A})+\hat{\delta}_{\hat{\lambda}}
\hat{\mathcal{A}}_{\mu}(\mathcal{A})=\hat{\mathcal{A}}_{\mu}(\mathcal{A}
+\delta_{\lambda}\mathcal{A}),
\end{eqnarray}
allows to express noncommutative gauge fields $\hat{\mathcal{A}}$ in terms of ordinary gauge fields $\mathcal{A}$ as
\begin{eqnarray}
\label{Aswef}
\hat{\mathcal{A}}_{\mu}(\mathcal{A})&=& \mathcal{A}_{\mu}-\frac{i}{4}\theta^{\rho\sigma}
\{ \mathcal{A}_{\rho},\partial_{\sigma}\mathcal{A}_{\mu}+\mathcal{F}_{\sigma\mu}
\}+\mathcal{O}(\theta^2),
\end{eqnarray}
where $\theta^{\rho\sigma}$ are noncommutativity parameters of the canonical commutation relation \eqref{moyal_nc}.
Here, our chosen commutation relation for the polar coordinates is $[\hat{r},\hat{\phi}]= i \theta  \hat{r}^{-1}$,
which is not in the canonical form. However, one can easily show that this
is exactly equivalent to $ [\hat{r}^2,\hat{\phi}]=2i\theta$.
Thus for computational convenience, we use the commutation relation $[\hat{R}, \hat{\phi}]= 2 i \theta$
with $\hat{R}\equiv \hat{r}^2$.
The Seiberg-Witten map \eqref{Aswef} of
the noncommutative gauge fields, $\mathcal{\hat{A}}^{\pm}:=\hat{A}_{\mu}^{a\pm}\tau_{a}+\hat{B}_{\mu}^{\pm}\tau_{3}$,
 yields the following.
\begin{eqnarray}
\label{ncgauges}
\mathcal{\hat{A}}^{\pm}_{\mu}=
 \left(A_{\mu}^{a\pm}-\frac{\theta}{2}B_{\phi}^{\pm}\partial_{R}A_{\mu}^{a\pm}\right)\tau_{a}
+B_{\mu}^{\pm}\tau_{3}+\mathcal{O}(\theta^2).
\end{eqnarray}
Setting the two $U(1)$ fluxes as $B_{\mu}^{\pm}=B d\phi ~$ with constant $B$,
we obtain the following expressions for the noncommutative triad and spin connection:
\begin{eqnarray}
\label{nctriad}
\hat{e}^{0}&=& \left(m-\frac{\theta B}{2}m'\right)\left(\frac{r_{+}}{l}dt-r_{-}d\phi\right)+\mathcal{O}(\theta^2),
\nonumber \\
\hat{e}^{1}&=& l \left[\frac{m'}{n}-\frac{\theta B}{2}\left(\frac{m'}{n}\right)'\right]dR+\mathcal{O}(\theta^2),
\nonumber \\
\hat{e}^{2}&=& \left(n-\frac{\theta B}{2}n'\right)\left(r_{+}d\phi-\frac{r_{-}}{l}dt\right)+\mathcal{O}(\theta^2),
\\
\label{ncspinc}
\hat{\omega}^{0}&=& -\frac{1}{l}\left(m-\frac{\theta B}{2}m'\right) \left(r_{+}d\phi-\frac{r_{-}}{l}\right)+\mathcal{O}(\theta^2),
\nonumber \\
\hat{\omega}^{1}&=&\mathcal{O}(\theta^2),
\nonumber \\
\hat{\omega}^{2}&=&-\frac{1}{l} \left(n-\frac{\theta B}{2}n'\right)
\left( \frac{r_{+}}{l}dt-r_{-}d\phi\right)+\mathcal{O}(\theta^2), \nonumber
\end{eqnarray}
where ${}'$ denotes the differentiation with respect to $R=r^2$.
We now define the metric in the noncommutative case as
\begin{equation}
\label{ncmetric}
d\hat{s}^2 :=\hat{g}_{\mu\nu}dx^\mu dx^\nu \equiv \eta_{ab}\hat{e}_{\mu}^{a}\star \hat{e}_{\nu}^{b}dx^{\mu}dx^{\nu} ,
\end{equation}
and with this definition we get a real metric($\hat{e}_{\mu} \star \hat{e}_{\nu}=\hat{e}_{\mu}\hat{e}_{\nu}$).
Rewriting $R$ in terms of $r$, we now get
\begin{eqnarray}
\label{ncmetricpol}
d\hat{s}^2=-f^2dt^2+\hat{N}^{-2}dr^2+2r^2 N^{\phi}dtd\phi
+\left(r^2+\frac{\theta B}{2}\right)d\phi^2+\mathcal{O}(\theta^2),
\end{eqnarray}
where $ N^{\phi}=-\frac{r_{+}r_{-}}{lr^2}$, $f^2=\frac{(r^2-r_{+}^2-r_{-}^2)}{l^2}-\frac{\theta B}{2l^2}$,
 $  \hat{N}^2=\frac{1}{l^2 r^2}\left[ (r^2-r_{+}^2)(r^2-r_{-}^2)
-\frac{\theta B}{2}\left(2r^2-r_{+}^2-r_{-}^2\right) \right]$.
In this solution, the apparent horizon (denoted as $\hat{r}$) which is determined by
\begin{eqnarray}
\label{ahor}
\hat{g}^{rr}=\hat{g}_{rr}^{-1}=\hat{N}^2=0,
\end{eqnarray}
and the Killing horizon (denoted as $\tilde{r}$) which is determined by
\begin{equation}
\label{khor}
 \hat{g}_{tt}-\hat{g}_{t\phi}^2/\hat{g}_{\phi\phi}=0,
\end{equation}
 are given as follows:
\begin{eqnarray}
\label{apparenthp}
\hat{r}_{\pm}^{2}&=&r_{\pm}^{2}+\frac{\theta B}{2}+\mathcal{O}(\theta^2),\\
\label{killinghp}
\tilde{r}_{\pm}^2&=&r_{\pm}^2 \pm \frac{\theta B}{2}
\left(\frac{r_{+}^2+r_{-}^2}{r_{+}^2-r_{-}^2}\right)+\mathcal{O}(\theta^2).
\end{eqnarray}
In the classical case, the apparent and Killing horizons coincide for stationary black holes.
Note that here the apparent and Killing horizons do not coincide.
Only in the non-rotating limit in which the classical inner horizon vanishes, $r_{-}=0$,
  we see that the two outer horizons coincide as in the classical case.

\section{Noncommutative BTZ solution in rectangular coordinates}
\label{secBTZ}

We now consider the noncommutative BTZ solution in the rectangular coordinates
following \cite{ell02}.
In order to evaluate the Seiberg-Witten map in the rectangular coordinates
we first have to express the classical solution of the $U(1,1) \times U(1,1)$ gauge fields
 in terms of the rectangular coordinates.
We again set the two $U(1)$ fluxes as $B_{\mu}^{\pm}=B d\phi=B(xdy-ydx)/r^2$ with constant $B$,
and the classical $SU(1,1) \times SU(1,1)$ solution in the rectangular coordinates is given by
\begin{eqnarray}
A^{0\pm} &=& \pm \frac{m(r_{+}\pm r_{-})}{l^2}\left[dt \pm \frac{l}{r^2}(ydx-xdy)\right],
\nonumber \\
A^{1\pm}&=& \pm \frac{1}{\sqrt{(r^2-r_{+}^2)(r^2-r_{-}^2)}}(xdx+ydy),
\\
A^{2\pm} &=& -\frac{n(r_{+}\pm r_{-})}{l^2}\left[dt \pm \frac{l}{r^2}(ydx-xdy)\right].\nonumber
\end{eqnarray}

Now performing the Seiberg-Witten map as in the previous section,
and using the relations ~ $\hat e/l=\hat{\mathcal{A}}^{+}+\hat{\mathcal{A}}^{-}$
and $ \hat \omega=\hat{\mathcal{A}}^{+}-\hat{\mathcal{A}}^{-}$,
we obtain the noncommutative triad and spin connection in the rectangular coordinates
up to first order in $\theta$ as follows.
\begin{eqnarray}
\label{nctrirectang}
\hat{e}^{0} &=&\frac{r_{+}[r^2-r_{+}^2-\theta B/4]}{l\sqrt{(r^2-r_{+}^2)(r_{+}^2-r_{-}^2)}}dt
+\frac{r_{-}}{r^2}\sqrt{\frac{r^2-r_{+}^2}{r_{+}^2-r_{-}^2}}
\left[ 1+\frac{\theta B}{4r^2}
\left(\frac{r^2-2r_{+}^2}{r^2-r_{+}^2}\right)\right](y dx-x dy),
\nonumber \\
\hat{e}^{1} &=& -\frac{l(r^2+r_{-}^2)}{(r^2-r_{+}^2)(r^2-r_{-}^2)}
\left[ 1-\frac{\theta B}{4r^2}
\frac{r_{+}^4(r^2-2r_{-}^2)-r_{-}^4(r^2-2r_{+}^2)}{(r_{+}^2-r_{-}^2)
(r^2+r_{-}^2)\sqrt{(r^2-r_{+}^2)(r^2-r_{-}^2)}}\right] (xdx+ydy),
\nonumber \\
\hat{e}^{2} &=& \frac{r_{-}[r^2-r_{-}^2-\theta B/4]}{l\sqrt{(r^2-r_{-}^2)(r_{+}^2-r_{-}^2)}}dt
-\frac{r_{+}}{r^2}\sqrt{\frac{r^2-r_{-}^2}{r_{+}^2-r_{-}^2}}
\left[ 1+\frac{\theta B}{4r^2}
\left(\frac{r^2-2r_{-}^2}{r^2-r_{-}^2}\right)\right](y dx-x dy),
\\
\label{ncspincrectang}
\hat{\omega}^{0} &=&\frac{r_{-}[r^2-r_{+}^2-\theta B/4]}{l^2\sqrt{(r^2-r_{+}^2)(r_{+}^2-r_{-}^2)}}dt
+\frac{r_{+}}{lr^2}\sqrt{\frac{r^2-r_{+}^2}{r_{+}^2-r_{-}^2}}
\left[ 1+\frac{\theta B}{4r^2}
\left(\frac{r^2-2r_{+}^2}{r^2-r_{+}^2}\right)\right](y dx-x dy),
\nonumber \\
\hat{\omega}^{1} &=& 0,
\nonumber \\
\hat{\omega}^{2} &=& -\frac{r_{+}[r^2-r_{-}^2-\theta B/4]}{l^2\sqrt{(r^2-r_{+}^2)(r_{+}^2-r_{-}^2)}}dt
-\frac{r_{-}}{lr^2}\sqrt{\frac{r^2-r_{-}^2}{r_{+}^2-r_{-}^2}}
\left[ 1+\frac{\theta B}{4r^2}
\left(\frac{r^2-2r_{-}^2}{r^2-r_{-}^2}\right)\right](y dx-x dy).\nonumber
\end{eqnarray}
Here we may define the metric by the same relation \eqref{ncmetric} as in the polar coordinates case.
However, this definition yields complex valued metric components.
Noting that the length element $d\hat{s}^2$ in (\ref{ncmetric}) has symmetric summation
after which  its value becomes real, we now define the metric by
 $\hat{G}_{\mu\nu} \equiv (\hat{g}_{\mu\nu}+\hat{g}_{\nu\mu})/2$
 as in \cite{pinzul06}.
Reexpress the rectangular coordinates back into the polar coordinates,
we then get
\begin{eqnarray}
d\hat{s}^2 &:=& \hat{G}_{\mu\nu}dx^{\mu}dx^{\nu}
\nonumber \\
&=& -\mathcal{F}^2 dt^2+\mathcal{\hat{N}}^{-2}dr^2
+2r^2 N^{\phi}\left(1+\frac{\theta B}{2r^2}\right)dt d\phi
+\left(r^2+\frac{\theta B }{2}\right)d\phi^2,
\end{eqnarray}
where
\begin{eqnarray*}
\mathcal{F}^2&=&\frac{(r^2-r_{+}^2-r_{-}^2)}{l^2}-\frac{\theta B}{2l^2}=f^2,
\\
\hat{\mathcal{N}}^2&=&\frac{1}{l^2 r^2}\left[ (r^2-r_{+}^2)(r^2-r_{-}^2)
-\frac{\theta B}{2r^2}\left( r_{+}^2(r^2-r_{-}^2)+r_{-}^2(r^2-r_{+}^2)
\right)\right].
\end{eqnarray*}

This solution now yields the apparent and Killing horizons which are
determined by the same relations \eqref{ahor} and \eqref{khor}, respectively,
 as in the previous section as follows.
\begin{eqnarray}
\label{apparenth}
\hat{r}_{\pm}^{2}&=&r_{\pm}^{2}+\frac{\theta B}{2}+\mathcal{O}(\theta^2),\\
\label{killingh}
\tilde{r}_{\pm}^2&=&r_{\pm}^2 + \frac{\theta B}{2}
+\mathcal{O}(\theta^2).
\end{eqnarray}
In this rectangular coordinates case, unlike the polar coordinates case,
the apparent and Killing horizons do coincide.
Although
the inner and outer horizons are shifted from the classical value
by the same amount $\theta B/2$ due to noncommutative effect of flux,
the feature that the apparent and Killing horizons coincide matches
with the classical result.

\section*{Acknowledgments}
This talk is based on the works done in collaboration with Daeho Lee and Youngone Lee.
This work was supported
by the Korea Science and Engineering Foundation(KOSEF) grant
funded by the Korea government(MEST), R01-2008-000-21026-0.



\begin{thebibliography}{99}

\bibitem{gromoyal}
J. E. Moyal, 
Proc. Cambridge Phil. Soc. \textbf{45} (1949) 99.

\bibitem{sw99} N. Seiberg and E. Witten,
JHEP \textbf{09} (1999) 032.


\bibitem{ns2005} N.~Seiberg,
 [hep-th/0601234].

\bibitem{bcgss} M. Banados, O. Chandia, N. Grandi, F. A. Schaposnik, and G. A. Silva,
Phys. Rev. \textbf{D 64} (2001) 084012.


\bibitem{ckmz} S. Cacciatori, D. Klemm, L. Martucci, and D. Zanon,
Phys. Lett. \textbf{B 536} (2002) 101.

\bibitem{cgm} S. Carlip, J. Gegenberg, and R. B. Mann,
Phys. Rev. \textbf{D 51} (1995) 6854.


\bibitem{btz} M. Banados, C. Teitelboim, and J. Zanelli,
Phys. Rev. Lett. \textbf{69} (1992) 1849.


\bibitem{ell01}
Ee~C.-Y., D.~Lee, and Y.~Lee,
  [arXiv:0808.2330].
\bibitem{ell02} Ee C.-Y., D. Lee, and Y. Lee,
[arXiv:0812.3507].

\bibitem{pinzul06} A. Pinzul and A. Stern,
Class. Quantum Grav. \textbf{23} (2006) 1009.



\end{thebibliography}
\end{document}